\documentclass[journal]{IEEEtran}

\usepackage{setspace}
\usepackage{cite,graphicx,amsmath,amssymb,url}
\usepackage{fancyhdr}

\usepackage{mdwmath}
\usepackage{mdwtab}
\usepackage{balance}
\usepackage{xcolor}
\usepackage{bm,amsthm,mathrsfs}
\usepackage{threeparttable}
\usepackage{algorithm}
\usepackage{algorithmic,algorithmic}
\usepackage{multirow}
\usepackage{flafter}
\usepackage{diagbox}

\usepackage{bbm,multicol,multirow,makecell}
\usepackage{array,epsfig}
\usepackage[font=small]{subfig}
\usepackage{stfloats}
\usepackage{lipsum}
\usepackage{caption}

\usepackage[T1]{fontenc}

\allowdisplaybreaks

\begin{document}

\vspace{-2.5em}
\title{Mobile Edge Generation-Enabled Digital Twin: Architecture Design and Research Opportunities}

\author{\normalsize {
Xiaoxia~Xu,~
Ruikang~Zhong,~
Xidong~Mu,~
Yuanwei~Liu,~\IEEEmembership{\normalsize Fellow,~IEEE,}
Kaibin~Huang,~\IEEEmembership{\normalsize Fellow,~IEEE}
\vspace{-2.5em}
}

\thanks{Xiaoxia~Xu and Ruikang~Zhong are with the School of Electronic Engineering and Computer Science, Queen Mary University of London, London E1 4NS, U.K.}
\thanks{Xidong Mu is with the Centre for Wireless Innovation (CWI), Queen's University Belfast, Belfast, BT3 9DT, U.K.}
\thanks{Yuanwei~Liu is with the School of Electronic Engineering and Computer Science, Queen Mary University of London, London E1 4NS, U.K., 
and is also with the Department of Electronic Engineering, Kyung Hee University, Yongin-si, Gyeonggi-do 17104, Korea.}
\thanks{Kaibin Huang is with the Department of Electrical and Electronic Engineering, The University of Hong Kong, Hong Kong, China.}

}

\date{\today}
 \maketitle

 \begin{abstract}
    A novel paradigm of mobile edge generation (MEG)-enabled digital twin (DT) is proposed, which enables distributed on-device generation at mobile edge networks for real-time DT applications. 
    First, an MEG-DT architecture is put forward to decentralize generative artificial intelligence (GAI) models onto edge servers (ESs) and user equipments (UEs), 
    which has the advantages of low latency, privacy preservation, and individual-level customization. 
    Then, various single-user and multi-user generation mechanisms are conceived for MEG-DT, which strike trade-offs between generation latency, hardware costs, and device coordination. 
    Furthermore, to perform efficient distributed generation, two operating protocols are explored for transmitting interpretable and latent features between ESs and UEs, namely sketch-based generation and seed-based generation, respectively. 
    Based on the proposed protocols, the convergence between MEG and DT are highlighted.
    Considering the seed-based image generation scenario, numerical case studies are provided to reveal the superiority of MEG-DT over centralized generation.
    Finally, promising applications and research opportunities are identified.
    Code is available at \url{https://github.com/xiaoxiaxusummer/MEG_DT}.
    
\end{abstract}

\begin{IEEEkeywords}
    Artificial intelligence generated content (AIGC), digital twin, mobile edge computing, {m}obile edge generation.
\end{IEEEkeywords}

\section{Introduction}
    
    
    With recent decade's booming developments in artificial intelligence (AI) and wireless networks, 
    digital twin (DT) has emerged as a promising approach for creating virtual replications of real-world physical objects and processes \cite{DT_Enabling_techniques}. 
    By revolutionizing the fashions of modelling, monitoring, analytics, and decision-making for intricate real-world systems, 
    DT is anticipated to transform diverse facets of human society, spanning industry, education, agriculture, healthcare, and beyond.  
    To maintain real-time evolving duplications of physical entities, DT requires ubiquitous connectivity, high-speed transmission, low latency, and intensive computing on mobile devices \cite{Realtime_DT,6G_DT}.
    To meet these challenges, a prospective solution is to integrate DT with the forthcoming sixth generation (6G) wireless networks, which leads to a win-win option.  
    On the one hand, 6G wireless networks enable real-time synchronized DTs by incorporating new communication technologies (e.g., millimetre-wave/terahertz communications and satellite communications), 
    enhanced infrastructures (e.g., integrated sensing and communications), 
    as well as cloud/edge intelligence \cite{6G_DT}. 
    On the other hand, DT provides powerful tools to realize precise modelling (e.g., channel modelling) and self-sustaining management of 6G wireless networks \cite{DT_Enabled_6G}.  

    While holding immense potentials, establishing and maintaining high-fidelity virtual twins suffer from significant manpower and economic costs, 
    particularly for real-time DT constructions. 
    Fortunately, benefiting from advances in generative AI (GAI) and large-scale Transformer models, 
    AI generated contents (AIGC) has unlocked new pathways to generate high-fidelity and vivid digital contents in virtual worlds \cite{AIGC_survey}. 
    The great success of GAI models, such as Generative Pre-training (GPT) \cite{GPT} and Stable Diffusion \cite{Stable_Diffusion}, 
    has spawned numerous popular AIGC applications in information industries, involving text generation, 
    text-to-image generation, image translation/synthesis, and video generation. 
    With the impressive capabilities to comprehend data patterns and to mimic artistic creations, 
    AIGC can automatically and seamlessly map real-world contents into virtual twins, 
    thus enabling smart content transformation and enhancement for DT \cite{AIGC_survey}. 
    However, the stunning generative performance relies on large-scale GAI models \cite{Stable_Diffusion, GPT}. 
    Specifically, the popular GPT-3 model requires  175 billions parameters \cite{GPT}. 
    Moreover, the number of parameters in the most recent GPT-4 model has grown to an unprecedented 1.7 trillion, which is nearly 1000 times larger than GPT-3. 
    Since deploying these large-scale GAI models on resource-limited mobile devices leads to an unaffordable hardware complexity \cite{LargeModel_Training}, 
    their applications in DT systems require powerful cloud services from distant datacenter.

    To reduce the access latency of remote cloud servers and achieve ubiquitous AIGC services, 
    mobile edge networks can be deployed for real-time DT systems, which offload generation tasks from resource-limited user equipments (UEs) to nearby edge servers (ESs). 
    However, as DT systems comprise massive devices with intensive generation requests, 
    it is time-consuming and privacy-sensitive to transmit huge-volume data for accessing centralized GAI models at ESs.
    Against this background, 
    the concept of mobile edge generation (MEG) has been recently proposed \cite{MEG}. 
    Specifically, MEG decentralizes GAI models into sub-models deployed on UEs and ESs,which provides a tailored edge AI solution for mobile AIGC. 
    As a result, distributed generation can be achieved by transmitting intermediate features of GAI sub-models between different hardware devices without exchanging the raw data. 
    In comparison to centralized generation, MEG can significantly reduce the latency and hardware costs for mobile UEs to access large-scale GAI models. 
    Therefore, it enables an innovative paradigm to exploit large-scale GAI models like GPT-4 and  DALL-E  in real-time DT systems. 
    However, to the best of our knowledge, the integration of MEG and DT has not been explored yet. 
    To fill in this gap, we propose a novel MEG-enabled DT (MEG-DT) architecture in this article, targeted at realizing efficient generation 
    on edge devices for DT systems. 
    Our main contributions can be summarized as follows.
    
    \begin{itemize}
        \item We propose a novel MEG-DT architecture, which decentralizes large-scale GAI models onto heterogeneous hardware devices at mobile edge networks. 
        The proposed MEG-DT can provide low-latency, privacy-preserving, and individual-level generation for real-time DT services, thus addressing the limitations of centralized generalization. 
        We discuss the key advantages and conceive single-user and multi-user generation mechanisms.
        \item We explore two operating protocols in MEG, namely sketch-based generation and seed-based generation.  
        To achieve efficient distributed generation, these protocols transmit interpretable and latent features between heterogeneous hardware devices, respectively. 
        We highlight that the convergence of MEG and DT can be realized, which empowers autonomous DT operation and self-optimized MEG configuration. 
        \item We identify suitable generation mechanisms and protocols for image, video, and interactive-scene generations in MEG-DT.  
        We further provide numerical case studies for seed-based image generation in MEG-DT, 
        which demonstrates that the proposed architecture achieves high-quality and noise-resistant generation compared to centralized generation even in low signal-to-noise (SNR) conditions, whilst significantly reducing overheads. 
    \end{itemize}

\section{MEG-DT: Architecture, Key Advantages, and Mechanisms}

In this section, we first propose a novel MEG-DT architecture. 
Then, 
we highlight key advantages of MEG-DT 
and present single-user and multi-user generation mechanisms.

\vspace{-0.8em}
\subsection{The Proposed MEG-DT Architecture}
\begin{figure}[!t]
    \vspace{-3em}
    \begin{center}
        \centering
        \includegraphics[width=0.37\textwidth]{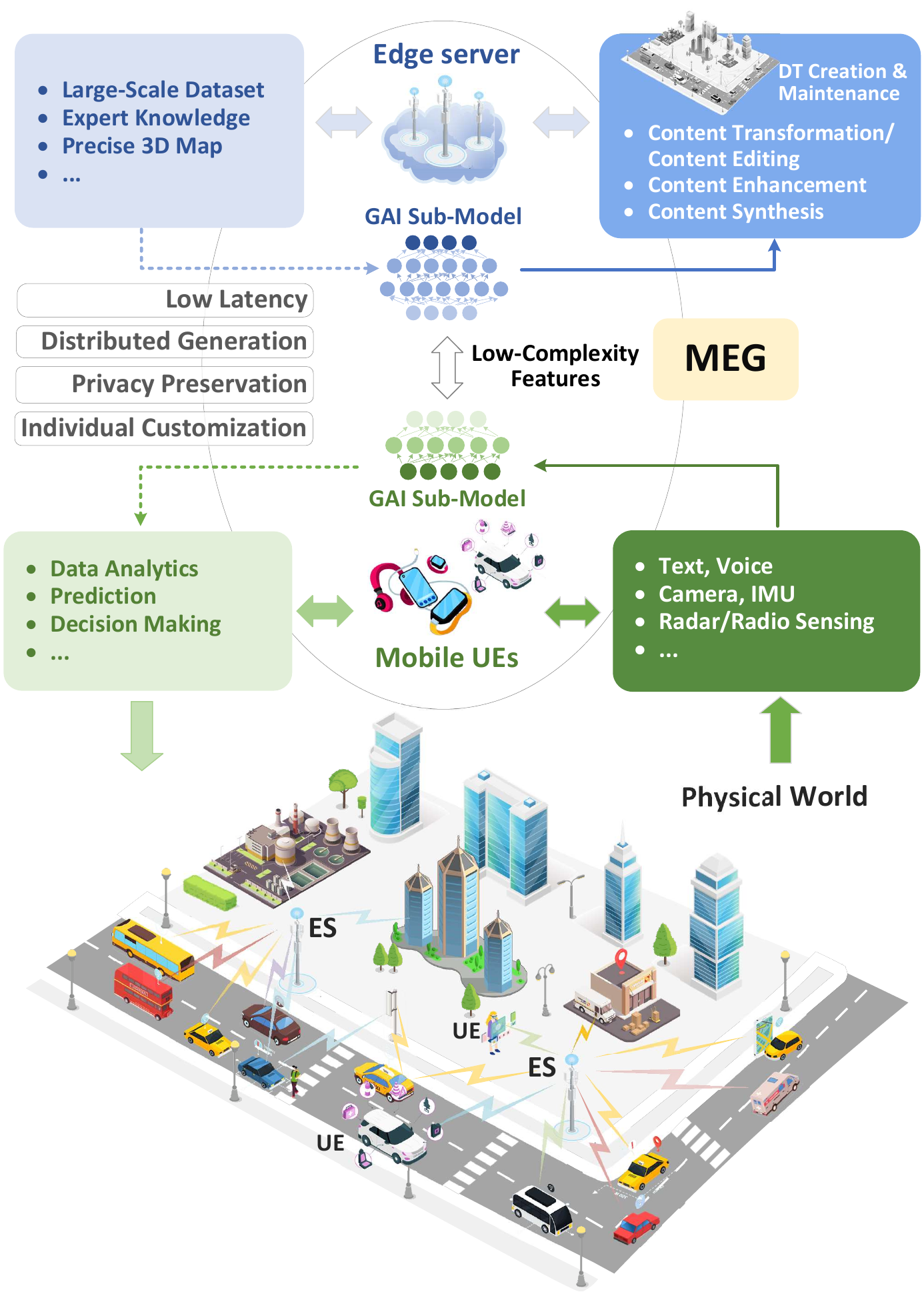}
        \vspace{-0.5em}
        \caption{The vision of the proposed MEG-DT architecture.}\label{fig_sys_model}
    \end{center}
    \vspace{-2.2em}
\end{figure}

As shown in Fig. \ref{fig_sys_model}, 
we consider a mobile edge network for MEG-DT that consists of heterogeneous hardware devices, including high-performance ESs and low-cost UEs. 
MEG provides mobile AIGC services across edge networks to construct and maintain DT.  
The key idea of MEG is to decompose the large-scale GAI models into decentralized GAI sub-models deployed on ESs and UEs, respectively. 
Instead of transmitting raw data, ESs and UEs only need to exchange low-dimension features extracted by GAI sub-models for content generation.   
Hence, distributed datasets, computing resources, and radio resources across mobile edge networks can be efficiently used.
Relying on the distributed generation, MEG-DT enables autonomous and cost-effective content transformation/editing, content enhancement, and content synthesis for virtual twins' creation and maintenance, 
as specified as follows. 
\begin{itemize}
    \item\textbf{Content Transformation and Editing}: MEG-DT can transform the physical-world data collected by mobile UEs' sensors  
    (e.g., voices, photos/videos, radar/LiDAR point clouds, and inertial measurement unit (IMU) records) into virtual replicas. 
    Thus, the digital mapping to real-world entities can be created automatically, e.g., for smart city DT creations. 
    The created virtual contents can be further edited by specific prompts according to users' intentions. 
    \item\textbf{Content Enhancement}: MEG-DT can improve the completeness and quality of virtual contents, 
    such as compensating for missing information, augmenting image/video quality, and mitigating distortion. 
    This can significantly enhance virtual replicas' fidelity and user experiences for immersive DT applications (e.g.,  cloud game and  extended reality (XR)). 
    \item\textbf{Content Synthesis}: MEG-DT can synthesize multi-modal virtual and/or reality data from different edge devices.  
    This enables the virtual-reality synchronization and the real-time interactions between virtual/reality entities for interactive DT applications (e.g., interconnected robotics and Metaverse). 
\end{itemize}

\vspace{-0.8em}
\subsection{Key Advantages}
Key advantages of MEG-DT can be summarized as follows.
\begin{itemize}
    \item \textbf{Low-Latency Feature Transmission:} MEG transmits features constructed from GAI sub-models, 
    which encode and compress the raw data into low-dimension representations.
    Compared to centralized generation that directly exchanges raw data between UEs and ESs, the communication overheads and end-to-end generation latency required by MEG-DT can be significantly reduced.
    \item \textbf{Efficient Edge Resource Utilization:} MEG enables distributed GAI model deployment and flexible collaborations between multiple UEs and ESs. 
    Therefore, the computing/radio resources and local datasets from heterogeneous devices can be efficiently orchestrated to enhance service efficiency 
    in various application scenarios.
    \item \textbf{Privacy and Security Preservation:} Unlike centralized generation that directly transmits users' data, 
        MEG-DT allows UEs/ESs to transmit implicit features that encodes users' private data to protect against eavesdropping.
        Therefore, MEG-DT can preserve the data privacy and improve system security for various DT applications.
    \item \textbf{Individual-Level Customized Generation}: Relying on MEG-DT, 
    the neural parameters and structures of UEs' local sub-models can be customized according to their individual preferences, personal generation purposes/demands, and particular hardware constraints.
\end{itemize}

\vspace{-0.8em}
\subsection{Single-User and Multi-User Generation for MEG-DT}\label{Sec_Generation_Mechanisms}
Based on the proposed MEG-DT architecture, we discuss single-user and multi-user generation mechanisms for different scenarios, 
which strike trade-offs between the latency, hardware complexity, and device coordination.

\begin{figure}[!t]
    \vspace{-1.8em}
    \begin{center}
        {\centering\scalebox{0.4}{\includegraphics{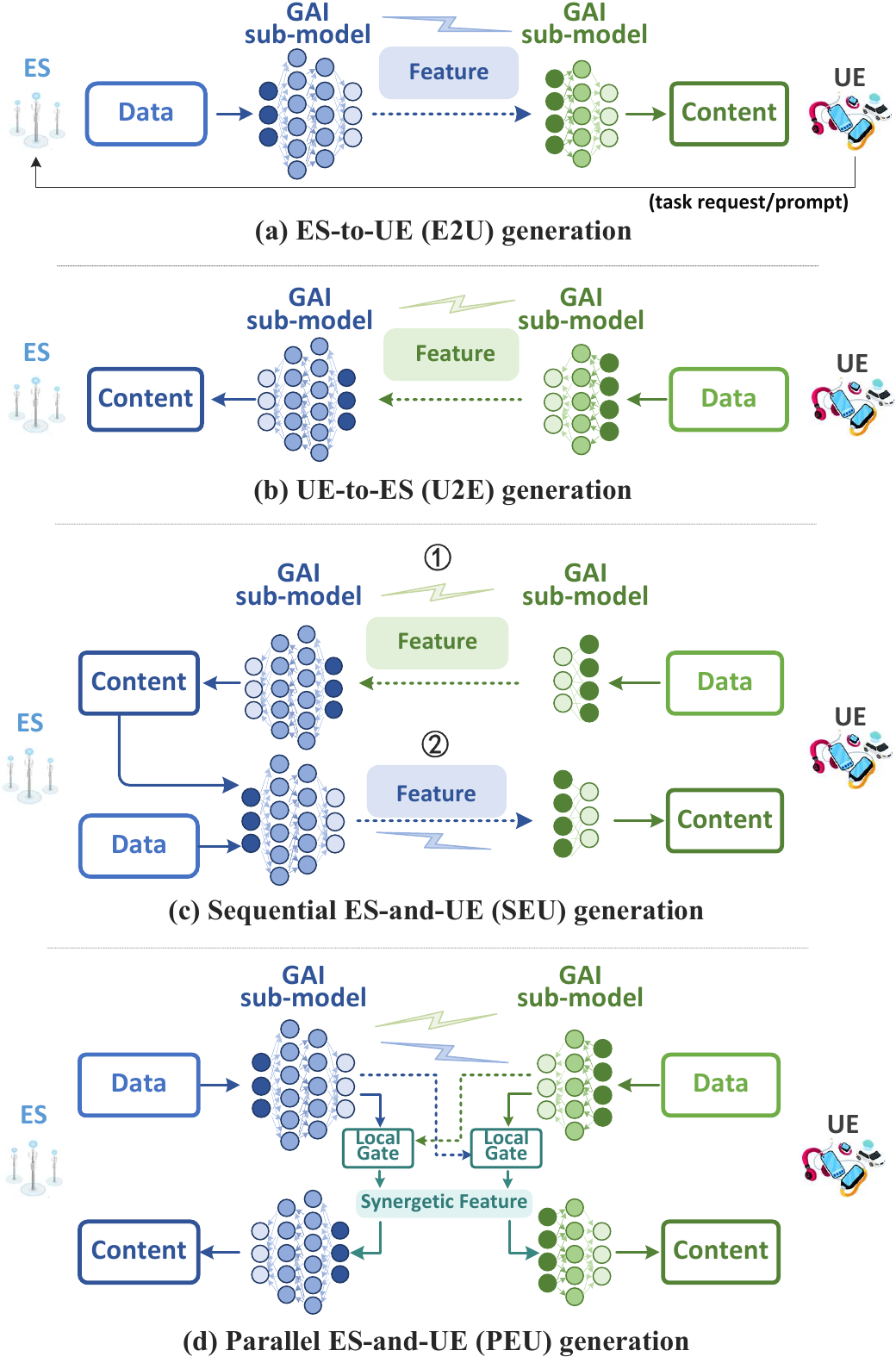}}} 
        \caption{Single-user generation mechanisms for MEG-DT.}\label{fig_protocol}
    \end{center}
  \vspace{-1.6em}
\end{figure}

\subsubsection{Single-User Generation}
As shown in Fig. \ref{fig_protocol}, the proposed MEG-DT architecture mainly contains four single-user generation mechanisms. 
Specifically, ES-to-UE (E2U) generation and UE-to-ES (U2E) generation performs feature encoding at the UE and the ES, respectively. 
Furthermore,  sequential ES-and-UE (SEU) generation and parallel ES-and-UE (PEU) generation enable feature encoding and content generation on both sides, as discussed as follows.

\begin{itemize}
    \item\textbf{E2U Generation:}
    The ES first receives uplink (UL) task requests from the UE and performs inference. 
    Then, low-dimension features are extracted from inference results and transmitted  to the UE via downlink (DL) channels.
    Using the received features, the UE further decodes desired contents. 
    This enables the UE to perform prompt-based generations by exploiting large-scale GAI sub-model and databases deployed at the ES.
\item\textbf{U2E Generation:} The UE encodes the collected sensing data into low-dimension features. 
By receiving features from UL transmissions, the ES generates digital contents.
This allows the ES to create and update virtual replicas exploiting the multi-modal sensing data from UE. 
\item\textbf{SEU Generation:} SEU generation allows the ES and the UE to sequentially perform feature encoding and transmissions for content editing/synthesis. 
The UE first encodes local data and uploads the encoded features to the ES by UL transmissions. 
Relying on this, the ES creates contents and transmits extracted features via DL channels to facilitate UE-side generation.
This mechanism can be employed to edit or synthesize contents when the UE has very limited computing and storage capacities. 
\item \textbf{PEU Generation:} 
PEU generation enables the UE and the ES to simultaneously transmit features 
in both UL and DL channels, and thus generate contents in parallel,  
as shown in Fig. \ref{fig_protocol}(d). 
A feature gate is learned to integrate features from both sides. 
Local copies of the feature gate are deployed at both the ES and the UE,
 thus enabling a shared view for synchronized and parallel generation.
Compared to SEU generation, PEU enables fast data synthesis and ES-UE interactions for delay-sensitive DT applications, but requires higher hardware costs for UEs.
\end{itemize}

\subsubsection{Multi-User Generation}
\begin{figure}[!t]
    \begin{center}\vspace{-2.3em}
        \centering
        \includegraphics[width=0.42\textwidth]{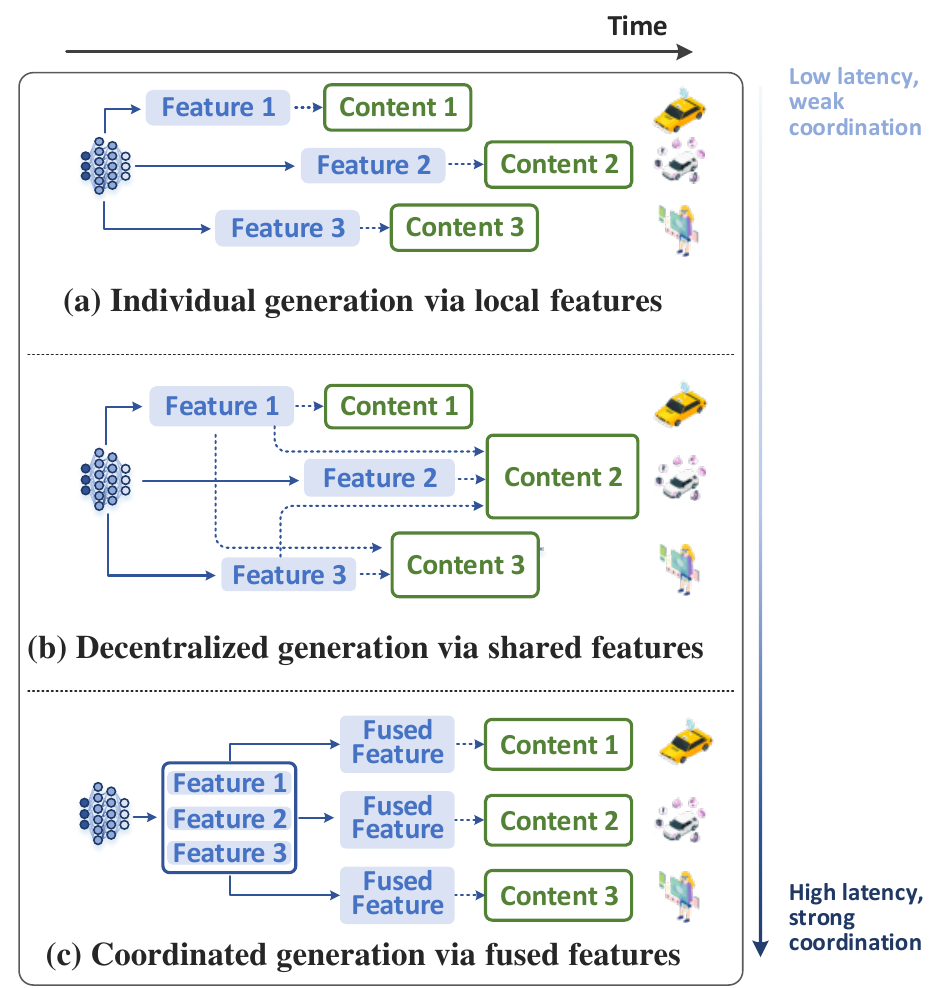}
        \caption{{Multi-user generation mechanisms for MEG-DT.}}\label{fig_MUprotocol}
    \end{center}
    \vspace{-1.9em}
\end{figure}

Maintaining real-time DTs requires devices' interaction/cooperation.
By integrating UEs' local features extracted from the surrounding environment, a global view of physical/digital world can be established for immersive experiences and cooperative decision-making. 
However, devices' interactions will introduce extra transmission latency.  
To strike a balance, we conceive three multi-user generation mechanisms, which can realize distinct levels of interaction/coordination and latency for MEG-DT, as shown in Fig. \ref{fig_MUprotocol}.   
We focus on UE-side cooperation and generation in this part. The ES-side cooperation can be achieved similarly.

\begin{itemize}
    \item\textbf{Individual Generation via Local Features:} 
    In the first level, UEs  independently perform single-user generation using their local features. As shown in Fig. \ref{fig_MUprotocol}(a), 
    each UE utilizes its sensors to collect local environmental information and individually interacts with the ES. 
    Thus, a real-time and local view of physical world can be acquired and projected onto the 3D map. 
    This empowers fast generation and decision-making for low-cost UEs.
    \item\textbf{Decentralized Generation via Shared Features:}
    The individual generation performance is limited by the local view of the real-world environment observed by each UE. 
    As a remedy, the second level envisions UEs to dynamically share their encoded features. 
    By integrating features shared by nearby UEs, each UE can asynchronously perform decentralized generation once the local observation is renewed, 
    thus enabling multi-device interactions with a low latency.
    \item\textbf{Coordinated Generation via Fused Features:}
    The third level enables UEs to thoroughly fuse their local features at the nearby ES for enhanced coordination.
    Exploiting the fused features, fully synchronized and coordinated multi-user generations can be realized. 
    This mechanism can be employed when UEs require large-scale data synthesis and mutual interactions, while tolerating increased latency and computing complexity.
\end{itemize}

\section{MEG-DT: Principle and Interplays}

To realize efficient distributed generation during the single-user and multi-user generations, a key issue is to effectively extract low-dimension features that encode local data of ESs/UEs.  
In this section, we will introduce the main principles to extract useful features by exploring two protocols, i.e., sketch-based generation and seed-based generation. 
Then, the convergence between MEG and DT will be discussed. 

\subsection{Protocol of MEG: Sketch-based or Seed-based?}
Depending on whether the transmitted features can be explicitly interpreted or not, 
we introduce two basic operating protocols for MEG, namely sketch-based generation and seed-based generation, 
as discussed as follows. 

\subsubsection{Sketch-Based Generation} 
Sketch typically refers to a simplified drawing that only retains object outlines and essential spatial characteristics of the original image. 
Derived from this concept in computer vision, 
we refer to sketch as a low-complexity but human-understandable representation for various types of data in DT.  
To name a few, sketches can represent summaries of texts,  contours of 3D maps, snapshots of videos, and so on.
Sketch-based generation compresses data into sketches for distributed generation. 
As such, users can gain a fast understanding of the generated contents before recovering the high-dimension data, 
thus easily examining and guiding creations during distributed generation.

\begin{figure}[!tb]
    \vspace{-2.3em}
    \begin{center}
        \centering
        \includegraphics[width=0.45\textwidth]{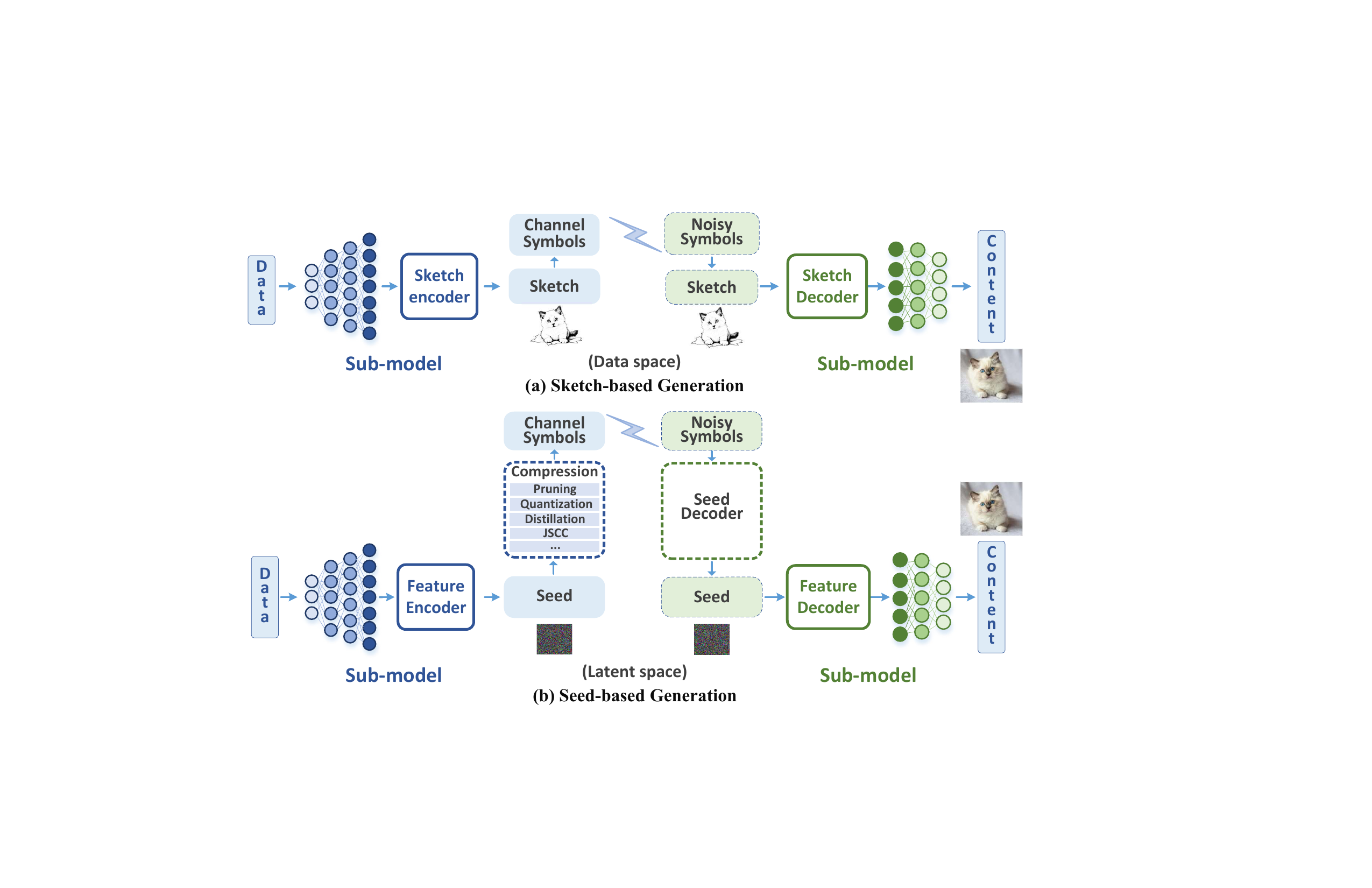}
        \caption{{Sketch-based and seed-based generation protocols.}}\label{fig_Generation_protocol}
    \end{center}
    \vspace{-2em}
\end{figure}

To efficiently perform sketch-based generation, ESs and UEs can jointly train domain-specific sketch encoder and decoder to bridge between the distributed GAI sub-models, as shown in Fig. \ref{fig_Generation_protocol}(a). 
Specifically, the encoder learns to extract a sketch of the raw data, thus preserving important information while reducing data volumes. 
Then, by transmitting the obtained sketch to the decoder deployed on another hardware equipment, the details will be reconstructed for high-quality generations. 
To efficiently compress and reconstruct data, the encoder and decoder can inherit the architectures from current GAI models for image-to-image transformation \cite{Image_Translation_Survey}, 
e.g., generative adversarial networks (GANs), variational auto-encoder (VAE), and diffusion models. 
These models can be further tailored for different generation purposes. 
During generation, text prompts can be combined to control the content editing and synthesis, thus flexibly synthesizing virtual/realistic datasets on ES and UEs. 

\subsubsection{Seed-Based Generation} 
Instead of transmitting sketches that have explicit interpretations, another promising approach is to transform the high-dimension data into latent features without intuitive explanations. 
These latent features are known as seeds. 
In comparison to the sketch-based generation that creates contents on the original feature space of the physical-world data, 
seed-based generation can deal with a more compact latent feature space. 
By identifying the underlying dependence of data implicitly, 
seed-based generation is more efficient to process large-scale or sequential data, such as time-varying sensing observations and videos. 

A simple implementation of seed-based generation is to directly split an existing GAI model (e.g., Stable Diffusion) into two sub-models, and then construct seeds by the intermediate hidden layers.
Similar to existing edge AI solutions, the edge-device co-inference can be achieved.
However, different from classification/regression tasks, MEG targets at content generation,  
where the end-to-end generation quality is sensitive to wireless channel noises and feature compression.
To fully exploit the potentials of seed-based generation and overcome these limitations,  
advanced MEG-DT strategies should be developed to realize efficient seed construction, compression, and source-channel coding,
thus enhancing on-device generation quality while relieving system complexity and overheads, as discussed as follows. 
\begin{itemize}
    \item\textbf{Seed and Sub-Model Compressions:} 
    To enable lightweight computing and low-latency generation, it is crucial to learn tiny GAI sub-models and acquire dimension-reduced seeds in MEG-DT. 
To fully reduce generation overheads exploiting latent features, both seeds and GAI sub-models  can be efficiently compressed. 
Specifically, the dimensions of seeds constructed by distributed sub-models should be adaptively lessened by removing the insignificant neurons and information via pruning. 
Furthermore, the GAI sub-models should be flexibly slimmed into tiny neural structures based on dynamic distillation \cite{Knowledge_Distiilation} or truncated diffusion, 
thus reducing computational burdens on UEs. 
    \item\textbf{End-to-End Coding and Transmission:} To enhance the end-to-end MEG performance, 
    GAI sub-models can be designed to accomplish joint source-channel coding (JSCC) for transmitting seeds in noisy wireless channels. 
    As shown in Fig. \ref{fig_Generation_protocol}(b), at seed transmitters, GAI sub-models learn to directly map the source signals of the constructed seeds into channel symbols.   
    Then, at seed receivers, seeds are further estimated from noisy channels to generate desired contents. 
    The distributed GAI sub-models can be jointly trained to enhance the resulting content generation quality, whilst minimizing seed distortions. 
    This can efficiently mitigate channel noises and potential hardware impairments for seed transmissions. 
\end{itemize}

\begin{figure*}[!t]
    \vspace{-2em}
    \begin{center}
        \centering
        \includegraphics[width=1\textwidth]{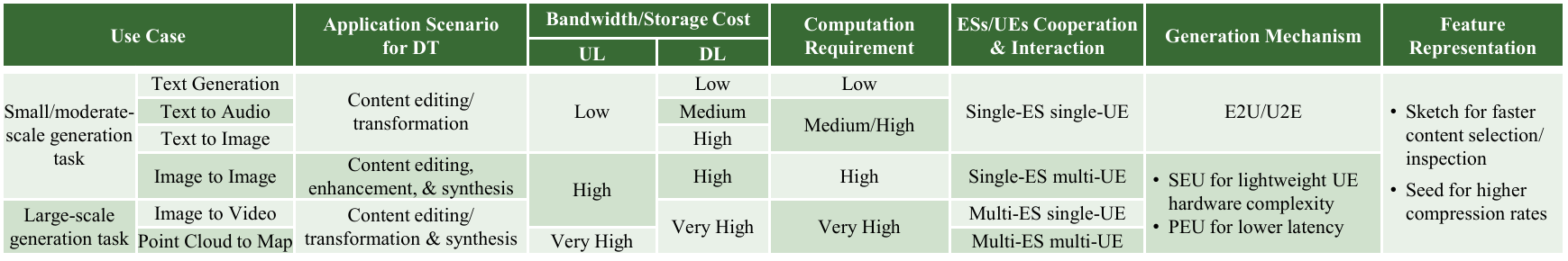}
        \caption{Suitable MEG mechanisms/protocols for various use cases and application scenarios.}\label{fig_tasks}
    \end{center}
    \vspace{-1.4em}
\end{figure*}

\vspace{-0.5em}
\subsection{Convergence between MEG and DT}
The proposed MEG-DT architecture empowers the convergence and interplays between MEG and DT, which can be analysed as follows.
\begin{itemize}
    \item \textbf{MEG Aided DT}:
    MEG enables autonomous creation, maintenance, and validation of DT on mobile devices.
    Specifically, the decentralized GAI sub-models can be trained to minimize virtual-reality divergence, 
    thus constructing high-fidelity digital replications. 
    Moreover, Vivid images, videos, and voices can be created by MEG by synthesizing inputs from mobile UEs based on multi-UE coordination. 
    Further, MEG can emulate expert behaviours in virtual worlds for decision-making. 
    Then, testing scenarios can be generated for validation, thus assisting the online self-prediction, self-diagnose, and self-management of DT systems. 
    \item \textbf{DT Aided MEG}: 
    DT can capture system dynamics of mobile edge networks based on data analytics and multi-physics simulation, 
    thus achieving self-optimized configuration of MEG. 
    For single-user generation, DT can empower low-overhead and high-reliability feature transmissions by simulating real-world channel environment and track UEs' mobility. 
    For multi-user/ES cooperative coordination, DT can provide predictive analytics of UEs' behaviours and environmental changes and derive 
    reliable and secure distributed computing schemes, which balances the computation/communication workloads and diagnoses potential malicious attacks for MEG.
\end{itemize}

\section{Case Studies}

This section discusses the applicability of distributed deployment and generation strategies of GAI models in various scenarios. 
Further, we provide numerical examples to demonstrate the performance of E2U seed-based image generation.

\vspace{-0.6em}
\subsection{Applicability Analyses}
DT involves the editing, transformation, and synthesis of multi-modal contents across texts, voices, images, videos, etc. 
The applicabilities of distributed deployment and generation strategies are summarized in Fig. \ref{fig_tasks}, as discussed as follows.
\begin{itemize}
\item For small-scale/modest-scale generation tasks (e.g., text creation and text-to-image generation), 
the distributed deployment across a single-ES single-UE system can satisfy computational requirements for 
popular GAI models, such as Stable Diffusion and DALL-E. 
The applicable generation mechanisms and feature representations can be analyzed as follows. 
\textbf{(i) Generation mechanisms:} Different generation mechanisms can be exploited for data compressions in DL/UL transmissions on demand. 
For UE-side text-to-image generations, since text uploading consumes much less resources than image downloading, E2U generation can be adopted to encode image features in DL transmissions. 
Similarly, U2E generation can be applied in the ES-side text-to-image generation.
For image-to-image and image-to-video generations, both UL and DL transmissions are bandwidth-thirsty. 
To reduce transmission overheads, we can exploit SEU generation to sequentially encode vision features at the UE and the ES. 
Moreover, PEU generation enables low-latency synthesis of images/videos/musics from the ES and the UE. 
\textbf{(ii) Feature representations:} For feature representations, sketch-based generation enables faster content selections and inspections. 
Before reconstructing high-dimension contents, users can quickly glance at sketches to filter out undesirable contents.
By contrast, the seed-based generation can achieve higher compression rates especially for video streams and preserve data privacy. 
\item For large-scale generation tasks 
(such as ultra-high-resolution (UHD) images/videos generation and metaverse applications),
the required GPU memory and computing power may overwhelm the capacities of a single-ES single-UE system. 
A tractable solution is to partition large-scale contents into multiple data slices that can be cooperatively generated by multiple interactive ESs/UEs. 
To update the local data slices at UEs, cooperative multi-user generation mechanisms (as discussed in Section \ref{Sec_Generation_Mechanisms}) 
can be utilized to enable low-cost synchronization and interactions. 
Moreover, to generate large-scale contents in parallel, different slices can be assigned to nearby ESs for cooperative multi-ES generation, thus fully utilizing edge computing resources.
\end{itemize}

\begin{figure*}[!t]
    \vspace{-2.2em}
      \centering
      \subfloat[\small{Scene generation quality}.]{\centering \scalebox{0.34}{\includegraphics{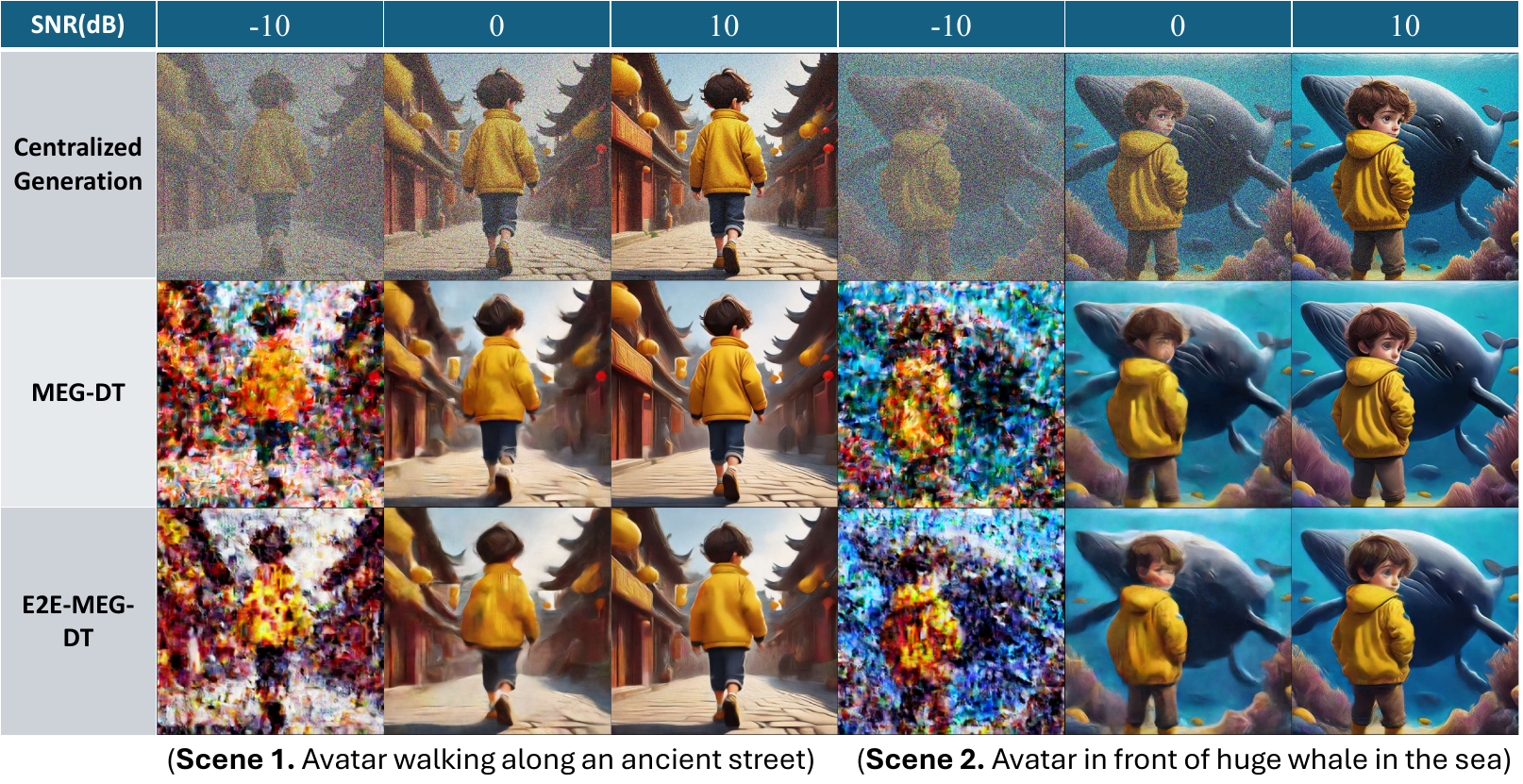}} }
      \quad
      \subfloat[\small{FID score}.]{\centering \scalebox{0.43}{\includegraphics{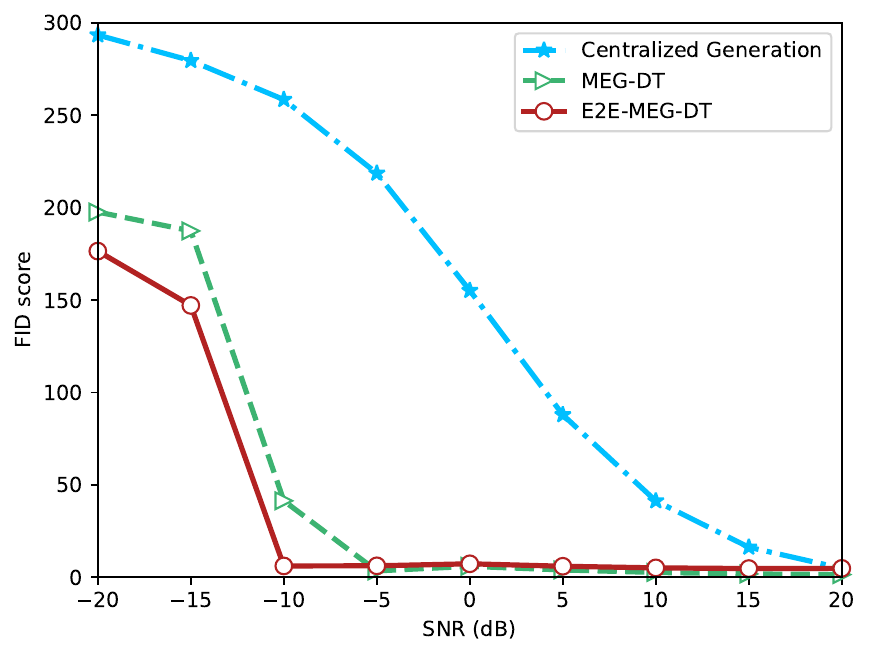}} }	
      \vspace{-0.2em}
      \caption{Performance comparisons of different schemes, where various virtual scenes are created for user's avatar in MEG-DT.}\label{fig_simulation}
      \vspace{-1.5em}
  \end{figure*}

\subsection{Performance Evaluation}
In this part, we provide case studies of text-to-image generation for the proposed MEG-DT architecture. 
We investigate the seed-based E2U generation under compact additive white Gaussian noise channels.
We evaluate the distortion between the reconstructed images and the reference images by Fréchet inception distance (FID) scores. 
Here, the reference images are generated by centralized GAI and are transmitted under perfect channel conditions. 
We assume $16$-bit floating point number and the average data transmission speed is $1$ Mbps.
The following three generation approaches are considered:

\begin{itemize}
\item \textbf{Centralized Generation:} The ES generates $1024\times 1024$ images leveraging a pre-trained Stable Diffusion XL (SDXL) model  \cite{Stable_Diffusion}, 
which are transmitted to the UE.
\item \textbf{MEG-DT:} The SDXL model is split and fine-tuned over noisy channels. 
The latent feature is sampled by the VAE encoder and U-Net denoiser deployed at the ES,
which is transmitted to the UE for image decoding.
\item \textbf{E2E-MEG-DT:} 
The latent features of MEG-DT are further compressed by merging similar neurons.
A pair of JSCC encoder/decoder is jointly trained with VAE decoder, which minimizes the end-to-end image distortions over noisy wireless channels. 
\end{itemize}

Fig. \ref{fig_simulation} presents the generation performance of different schemes to create virtual scenes for an avatar in DT. 
The transmission latencies for centralized generation, MEG-DT, and E2E-MEG-DT are $50.33$ seconds, $1.05$ seconds, and $0.58$ seconds, respectively. 
The corresponding end-to-end generation latencies are $57.91$ seconds, $8.63$ seconds, and $8.16$ seconds, 
where $12$ denoising steps are sampled by SDXL models for all schemes. 
Both MEG-DT and E2E-MEG-DT can achieve higher generation qualities than centralized generation whilst significantly reducing transmission overheads. 
For high SNR scenarios (i.e., SNR $\geqslant 0$ dB),  E2E-MEG-DT suffers from higher distortions than MEG-DT due to the additional losses caused by feature compressions. 
Nevertheless, for low SNR scenarios (i.e., SNR $< 0$ dB), E2E-MEG-DT scheme may outperform MEG-DT in both image generation qualities and compression rates,   
which demonstrates that the integrated JSCC encoder/decoder can improve noise resistance.

\section{MEG-DT: Applications and Open Problems}

\subsection{Promising Applications of MEG-DT}
We introduce several important applications of MEG-DT, which can be exemplified as follows.
\subsubsection{Smart City} Smart city involves densely deployed sensors, interconnected vehicles/robots, social networks, and cellular/WiFi networks. 
With the huge amounts of data continuously produced by these entities, virtual city twins can be constructed via MEG-DT to integrate observations of massive mobile devices. 
This will empower accurate time-series traffic forecasting, social event forecasting, and rational decision-making for city governors, traffic administrator, and network operators, etc. 
\subsubsection{Industry 4.0}: Industry 4.0 requires cyber-physical integration for smart manufacturing \cite{DT_Industry4}. 
MEG can easily build cyber-models of products using prompts and sensing data, thus achieving personalized and low-cost product designs. 
During manufacturing process, on-device virtual twins of smart factories can be created,  
thus supporting 24/7 real-time monitoring and quality control. 
\subsubsection{Metaverse and XR} MEG-DT can synthesize diverse datasets collected by different devices, 
thus realizing the convergence between virtual and real worlds to build XR and metaverse applications. 
For instance, the vehicular XR metaverse can be built by synthesizing massive traffic and driving data, thus improving robustness, driving safety, traffic control efficiency,
and sustainability \cite{GAI_XR_Metaverse}. 
\subsubsection{Healthcare} Enabling mobile AIGC in DT is promising to provide personalized healthcare \cite{DT_Healthcare}. 
 MEG-DT delivers a low-latency distributed solution for mobile AIGC, which can simulate data of  rare diseases, 
establish testbed for comprehensive drug validations and analytics, and provide ubiquitous personalized medical services. 

\vspace{-0.5em}
\subsection{Open Problems of MEG-DT}
While MEG-DT providing a promising distributed generation paradigm, there are still some open research problems in practical designs, which can be outlined as follows. 
\begin{itemize}
\item\textbf{Security}: 
Although seed-based generation enables privacy preservation, 
security remains an open challenge in MEG-DT especially for sensitive applications. 
Specifically, when sketches/seeds are attacked, corrupted, or modified by viruses, the content generation process may be collapsed. 
This poses an urgent need for developing advanced security techniques, e.g., blockchains, for trustworthy and invulnerable distributed generation.
Moreover, since sketches/seeds generally contain important object features and substantial semantic information, some privacy-sensitive information 
(e.g., local sensing data) may be eavesdropped by attackers through reverse engineering techniques. 
Hence, privacy-preserving techniques (e.g., differential privacy)  can be investigated, which protects privacy by adding controllable noises to data's statistical features. 
\item\textbf{Scalable Learning and Compression}:   
There are several candidate techniques for the compression of sketch/seed and GAI sub-models \cite{Compression}. 
Specifically, to achieve compact representations, inessential components can be removed by network sparsification methods (e.g., pruning), 
thus shrinking redundant neurons. 
As visual contents may be sensitive to pruning, another approach is to decompose the high-complexity tensors into low-rank and small-sized tensors via tensor decomposition, 
thus relieving memory and computational overheads. 
Instead of reducing model parameters, the quantization method focuses on achieving low-bit features, which are suitable for low-precision computing on mobile UEs. 
To further mitigate distortions for various application scenarios, knowledge distillation can flexibly distil tiny GAI sub-models by re-training over dedicated datasets.
In practice, MEG-DT suffers from both hardware and radio resource limitations, 
constituting an interest of developing scalable learning and compression techniques to balance between compression rate, hardware applicability, and generation quality. 
\item\textbf{Distributed Computing and Coordination}:
MEG-DT relies on efficient distributed computing and coordination. 
For individual generation, the distributed GAI sub-models need to be adaptively deployed and configured, 
thus enhancing the content fidelity while ensuring low latency and complexity. 
For coordinated generation, UEs should dynamically determine the optimal set of interacting devices and the optimal sketch/seed interaction frequency. 
Furthermore, to handle large-scale generation tasks, suitable data slicing and task offloading schemes are required for multi-ES cooperation. 
On the other hand, to accurately track environment changes whilst  refreshing digital replica  in the real time, 
the sensing, communication and computing resources over edge devices should be efficiently exploited. 
This requires the development of efficient coordination schemes between heterogeneous hardware devices and functionalities.
\item\textbf{Personalization and Customization}: 
Personalization is a crucial aspect for MEG-DT to meet users' individual expectations. 
For personalized generation, GAI sub-models need to be customized on individual UEs that experience different data distributions, performance metric requirements, and hardware/software limitations. 
The personalized customization can be mainly achieved via two approaches, i.e., data augmentation and model adaptation. 
On the one hand, data augmentation directly tackles the data distribution drift, 
which pre-trains the globally shared GAI sub-models with heterogeneous data and then fine-tunes the pre-trained sub-models via UE's local data.
On the other hand, model adaptation achieves enhanced and accelerated personalization of GAI sub-models via few-shot learning (e.g., meta-learning and transfer learning) or knowledge transfer techniques. 
Specifically, meta-learning learns an optimal GAI meta-model that can fast generalize across a range of tasks by maximizing the model parameter sensitivity. Based on the meta-model, personalized sub-models for particular device can be achieved via few-shot adaptation. 
Moreover, transfer learning focuses on diminishing the domain discrepancy between the globally shared sub-models and personalized sub-models.
Instead of transferring model parameters, knowledge transfer further shares knowledge between initial sub-models and personalized sub-models via soft predictions,  
which enables heterogeneous sub-model structures on different UEs.
\end{itemize}

\section{Conclusion}
A novel MEG-DT architecture has been proposed in this article for on-device DT generation at mobile edge networks. 
The key advantages of MEG-DT has been discussed, and single-user and multi-user generation mechanisms have been put forward. 
We introduced sketch-based and seed-based generations and discussed the convergence between MEG and DT. 
Furthermore, numerical case studies for the seed-based text-to-image generation have exhibited the superiority of MEG compared to centralized generation. 
The potential applications and open research problems have been identified.

\end{document}